\documentclass{iaus}
\usepackage{graphicx}

\title[
Variable Stars and Galactic Structure] 
{Variable Stars and Galactic Structure}

\author[Michael W. Feast \& Patricia A. Whitelock]   
{Michael Feast$^1$
 \and Patricia A. Whitelock$^2$}

\affiliation{$^1$ University of Cape Town and South African Astronomical Observatory
\\ email: {\tt mwf@ast.uct.ac.za} \\[\affilskip]
$^2$ South African Astronomical Observatory and University of Cape Town \\email: {\tt paw@saao.ac.za}}

\pubyear{2014}
\volume{298}  
\pagerange{XX -- YY}
\jname{IAUS\,298 Setting the scene for Gaia and LAMOST}
\editors{S. Feltzing, G. Zhao, N.\,A. Walton \& P.\,A. Whitelock, eds.}
\begin{document}

\maketitle

\begin{abstract}
Variable stars have a unique part to play in Galactic astronomy. Among the most important of these variables are the Cepheids (types I and II), the RR Lyraes and the Miras (O- and C-rich). The current status of the basic calibration of these stars in their roles as distance, structure and population indicators is outlined and 
some examples of recent applications of these stars to  Galactic
and extragalactic problems is reviewed. The expected impact of Gaia on this type of work is discussed and the need for complementary ground based observations, particularly
large scale near-infrared photometry, is stressed.  
\end{abstract}

\firstsection 
\section{Introduction}
Gaia will obtain parallaxes, proper motions, radial velocities and metallicities
([Fe/H]) for a vast number of stars and other objects of all types, and these
data will tell us much about our own Galaxy. One might ask why we should be 
particularly interested in the Gaia data for pulsating variable stars so far as Galactic composition, structure
and dynamics are concerned. In the present review we discuss the importance
of some major classes of pulsating variables, classical Cepheids, RR Lyrae
variables, type II Cepheids and Miras. These variables, once their luminosities
are calibrated by Gaia, will provide accurate distances both within our
Galaxy and beyond, reaching  beyond the distances achieved by Gaia itself. 
We will also point out some of the special problems that these stars pose for Gaia.

\section{Gaia parallaxes and their use}
Table 1 gives the expected accuracy ($\sigma$) of the absolute magnitudes determined for
single relatively bright stars ($6<V<12$\,mag) and assuming that for such stars 
$\sigma_{\pi} =10\mu as$ (see ESA Gaia web page). The table also gives an estimate of the
statistical bias expected in the absolute magnitude calibration (Lutz-Kelker type bias). These values of the bias are small. However, they are significant if one is
aiming to obtain distance scales with an accuracy of 2 percent or better (0.04\,mag).
Such accuracies are already being claimed for the  distance modulus of the LMC (0.04\,mag from eclipsing binaries (Pietrzy\'nski et al. 2013) and as small as 0.03\,mag  for classical Cepheids,
if metallicity effects are ignored). Furthermore, these bias corrections are
asymmetrical and have large uncertainties (Koen 1992) due to their statistical nature
and to the assumptions made in estimating them. Thus it is unlikely that stars
in a given sample are always selected on the basis of parallax alone or that
they are drawn from a population distributed uniformly in space. Note also 
that the bias correction required depends on the use to which the parallaxes are to be put (e.g. estimating distance moduli or estimating distance (see, e.g., Feast 2013). It is interesting to note that some VLBI parallaxes of star forming regions (e.g. Reid et al. 2009) have percentage accuracies similar to those expected from Gaia and are presumably subject to bias (Stepanishchev \& Bobylev 2013) though
the effect of this on derived Galactic constants is probably complex (see Bovy, these proceedings). These considerations are 
important for the use of Gaia parallaxes in calibrating luminosities. In the next few
sections we summarize the present status of the luminosity calibration of some well known variable stars and indicate what advances we can expect from Gaia.

\begin{table}
  \begin{center}
  \caption{Accuracy of Gaia parallaxes.}
  \label{NNN:tab1}  
 {\scriptsize
  \begin{tabular}{cccc}\hline 
r (kpc) & {\bf $\sigma_{\pi}/\pi$} & {\bf $\sigma$} (mag) & bias (mag)\\ 
\hline
1 & 0.01 & 0.022 \\
2 & 0.02 & 0.043 & $\sim$0.01\\
4 & 0.04 & 0.085 \\
5 & 0.05 & 0.11 & $\sim$0.02\\
  \end{tabular}
  }
 \end{center}
\vspace{1mm}
 
\end{table}

\section{Classical Cepheids}

\begin{figure}
\begin{center}
 \includegraphics[width=2.5in]{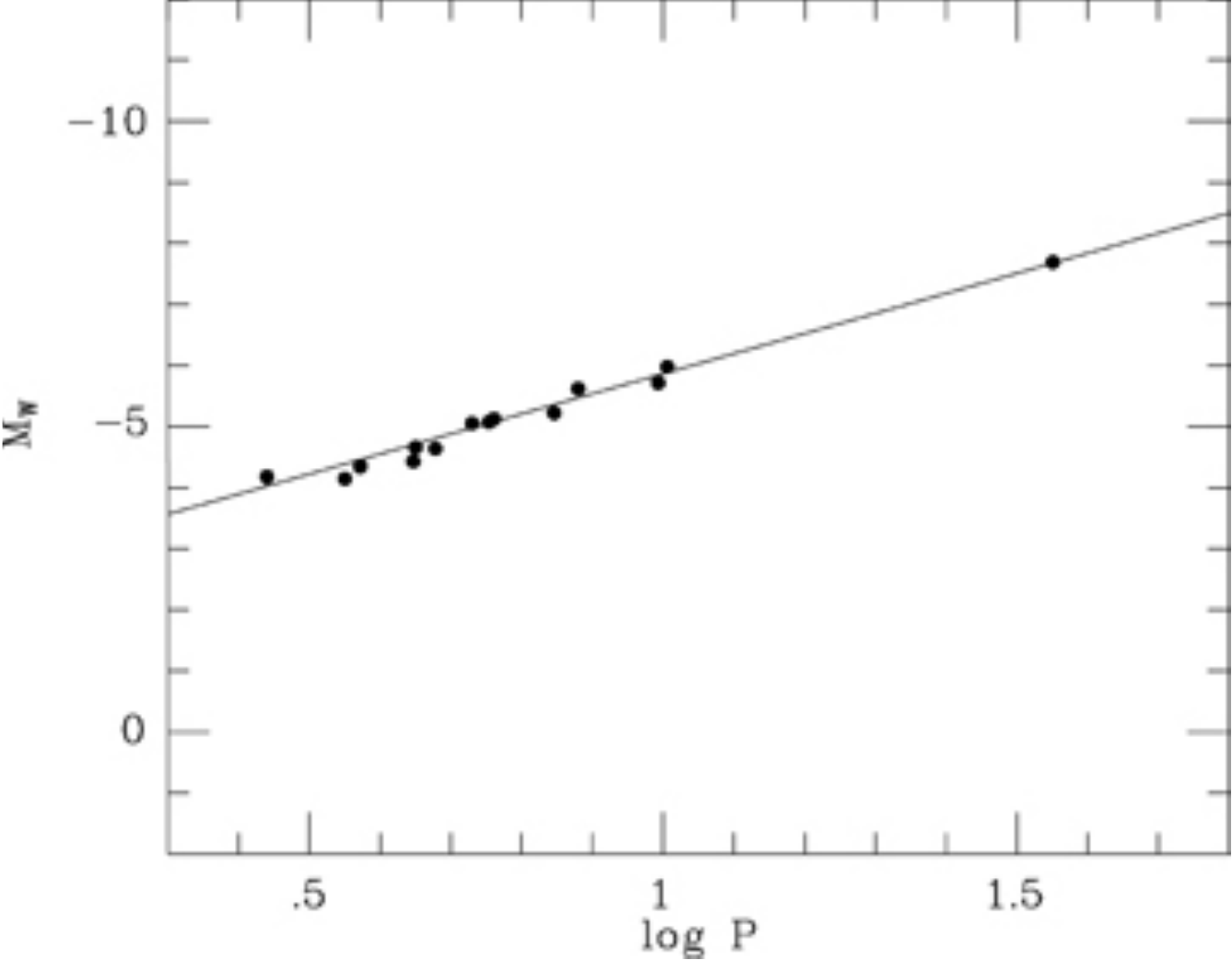} 
 \caption{Reddening-free PLC relation for Cepheids with good HST and Hipparcos Parallaxes (Van Leeuwen et al.\ 2007).}
   \label{Cepheid_PLC}
\end{center}
\end{figure}

 Fig.\,\ref{Cepheid_PLC} shows a reddening-free period-luminosity-colour (PLC) relation in $V,I$
($PW_{VI}$) for classical Cepheids with the best HST and Hipparcos parallaxes
(Benedict et al.\ 2007, van Leeuwen et al.\ 2007). The slope of this relation 
is close to that found for LMC Cepheids, which are metal deficient with respect to those in the solar neighbourhood. However, the slope evidently depends rather
critically on one point. Also, both in the Galaxy and beyond, the long period 
Cepheids, being bright, are important.
If it is assumed that the Galactic relation has the same slope as that found in the LMC, then a Galactic zero-point can be found from 
HST and Hipparcos parallaxes using the method of reduced parallaxes. This avoids
some of the bias problems just discussed  (see e.g. Feast 2002, 2013). In this way a zero point with a standard error of 0.03\,mag is obtained. This assumes the LMC slope.
The effects of metallicity on this slope and the zero point are at present poorly understood
though they are probably small.
 
How will Gaia help clarify this situation?   
   
For a sample of 32 LMC Cepheids with $1.0<\log P<1.7$, reddening-free 
$W_{VI} - \log P$ and $W_{JK_{S}}-\log P$ relations show a dispersion (per star)
of $\sigma = 0.175$\,mag and 0.085\,mag respectively (Feast et al. 2012). The scatter may have a component due to the thickness of the LMC, but is probably mainly observational: that in $VI$ being greater due to the coefficient of the colour term in ${VI}$ being numerically
larger. In using Gaia data on Cepheids to calibrate the $PW$ relations
it is obvious that one must aim to keep bias effects very small and it is also
desirable that the individual uncertainties are if possible considerably smaller
than the LMC scatter just quoted. As a rough guide Table 1 suggests that one should be limiting oneself to Cepheids within about 2\,kpc. With this restriction
we can expect many Cepheids with periods less that 10 days with 
$\sigma_{\pi}/\pi \lesssim 0.02$, perhaps about 20 with the same uncertainty
and periods over 10 days, and rather few in this category with periods over 20 days.
Of course, all this may change when the actual performance of Gaia is known, but
it suggests that at the shorter periods there will be sufficient data to study the
various Cepheid relations in some detail, including estimates of their intrinsic
scatter. At the longest periods we may be forced to employ the method of reduced parallaxes on more distance samples, possibly grouping the stars according  to period.

\begin{figure}
\begin{center}
 \includegraphics[width=3.3in]{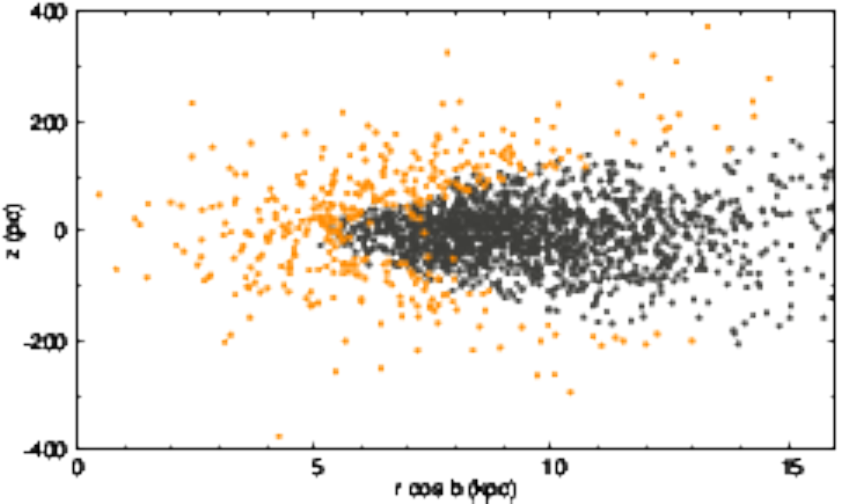} 
 \caption{The simulated inner Galaxy ($|l| < 5^{\circ}$), with projected distance to the Sun plotted versus height above the Galactic plane. The light dots correspond to Cepheids that will be observable by Gaia ($G < 20$) and the dark dots correspond to Cepheids that are too faint to be observed by Gaia ($G > 20$) (Windmark et al.\ 2011).}
   \label{Windmark}
\end{center}
\end{figure}

      The scatter in the LMC $W_{JK_S} - \log P$ relation given above is obviously
a maximum value for the intrinsic scatter. This being the case Table 1 suggests
that beyond about 4\,kpc this relation (calibrated by Gaia) will provide better distances for individual Cepheids than the Gaia parallaxes themselves, again provided the metallicity effects are negligible or can be quantified. This is important since
near-infrared ($JHK_S$) observations of Cepheids have, potentially, a much
wider reach over our Galaxy than Gaia itself. Fig.\,\ref{Windmark} illustrates this point. It shows a likely distribution of Cepheids in a direction toward the Galactic centre. 
The dark points are Cepheids which will not be observable by Gaia, primarily due to interstellar extinction. However, they will be
well within the range of current near-infrared facilities. Note that the 20 day
Cepheids found by Matsunaga et al.\ (2011) in the Galactic centre have $A_{V} \sim 25$\,mag but are still quite
bright in the infrared ($K_S \sim 10$\,mag). This highlights the importance of ground-based
near-infrared photometry of Cepheids. This together with Gaia results on
the $W_{JK_S}-\log P$ zero-point, intrinsic scatter and (hopefully) metallicity
effects, will allow the distribution (and with (infrared) velocities) the kinematics
of Cepheids over much of the Galactic disc to be determined. Such studies can be made as a function
of age,  
since Cepheid age is related to period:
$\sim 25$\,Myr at 20 days, $\sim 65$\,Myr at 6 days (Bono et al.\ 2005).

\section{RR Lyrae Variables}
The RR Lyrae variables are indispensable tracers of, and distance indicators for,
old stellar populations in our own and nearby galaxies. They are of current
importance in two major areas in our own Galaxy: the distances and hence the ages
of globular clusters, and the structure and kinematics of the Galactic halo
and its implication for galaxy formation. Two areas of research on the Galactic
halo are, first, the nature, extent and significance of infalling streams, and
second, the general overall structure of the halo. As regards the latter point,
there is a current debate as to whether or not the halo consists of two major
components: an inner, prograde, halo and an outer, retrograde, halo. 
The evidence for a two component model has been summarized by Carollo et al.\ (2007, 2010) and by Beers et al. (2012), but it has been criticized by Sch\"onrich et al.\ (2011) primarily on the basis of the distances adopted by Carollo/Beers.

\begin{figure}
\begin{center}
 \includegraphics[width=2.6in]{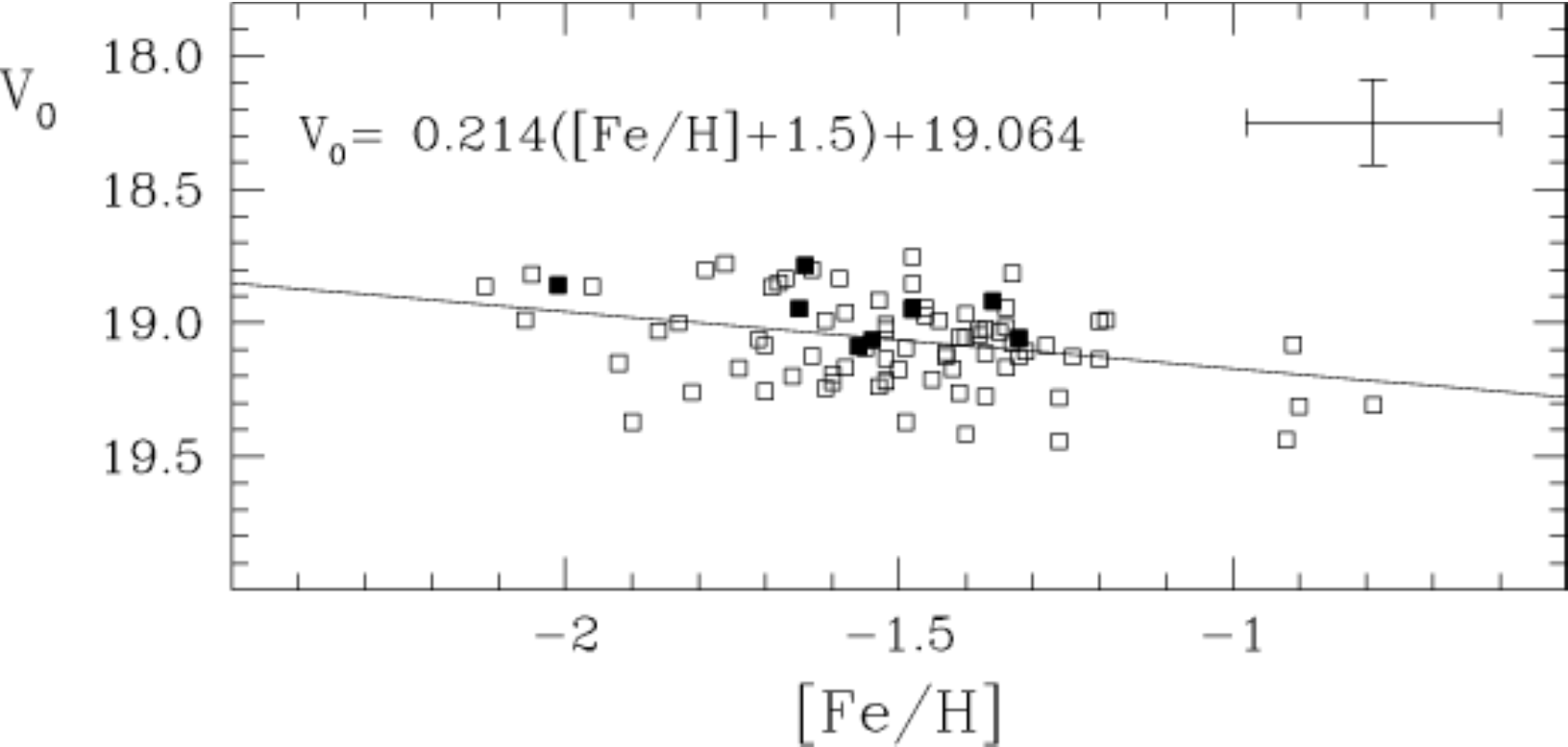} 
 \caption{Reddening corrected mean magnitude of the RR Lyrae variables in two LMC fields as a function of the metallicity [Fe/H]. Filled symbols are double-mode pulsators; open symbols ab- and c-type RR Lyrae stars (Gratton et al.\ 2004).}
   \label{Gratton}
\end{center}
\end{figure}

\begin{figure}
\begin{center}
 \includegraphics[width=2.8in]{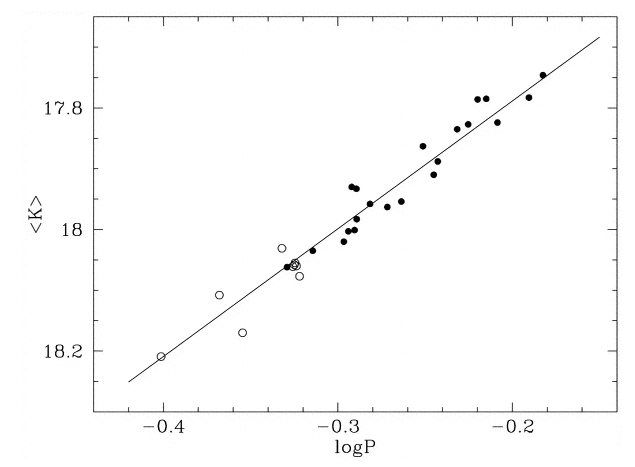} 
 \caption{The PL($K$) relation for the Reticulum RR Lyrae stars (Dall'Ora et al.\ 2004).}
   \label{Dallora}
\end{center}
\end{figure}

 There are two favoured ways of using RR Lyraes for distance determination.
 The first is a $V-[Fe/H]$ relation. Perhaps the best example of this is from
the LMC (Fig.\,\ref{Gratton}). As can be seen the scatter
is quite large. Much of this is probably due to the depth of the LMC halo
though uncertainties in reddening corrections and in [Fe/H] values may contribute.
At present the intrinsic width of the relation is not well determined. The second is an
$M_{K_S} -\log P$ relation. This is very promising as can be seen from
Fig.\,\ref{Dallora},  which shows the relation for the LMC globular cluster Reticulum
 (Dall'Ora et al.\ 2004). The scatter per star is very low (0.03\,mag). 
The dependence of the zero-point of this relation on [Fe/H] is not well determined,
but may well be quite small (see Feast 2013 for a discussion).

The current calibration of the two relations from the HST parallaxes of five
nearby RR Lyraes (Benedict et al.\ 2011) gives the following zero-points:\\
$ \hfill M_{V} =0.46 \pm 0.03 \ {\rm mag \ at} \ [Fe/H] = -1.5,  \hfill $\\
$  M_{K_S} = -0.54 \pm 0.03\ {\rm mag  \  at} \ \log P =-0.28. $\\
Whilst this calibration seems rather secure and, for instance, leads
to a distance modulus near 18.5\,mag for the LMC in agreement with several other
indicators, statistical parallax work gives
absolute magnitudes which are about a third of a magnitude fainter.
The differences ($\Delta$) are:\\
$\Delta M_{V} = 0.33 \pm 0.13$mag (Gould \& Popowski (1998) and earlier papers),\\
$\Delta M_{K} = 0.37 \pm 0.08$mag (Dambis 2009).\\
This is an 18 percent difference in the distance scale. If the statistical parallax
result were used it would imply an LMC modulus near 18.2\,mag, and a distance
to the Galactic centre near 7\,kpc, in contrast to most other estimates which are near 8\,kpc (see, e.g. Feast (2013) for a discussion of LMC and Galactic centre distances). This difference is not understood at present. There seems a possibility that the statistical parallax result might be affected by the relatively simple model of the halo assumed. 

Clearly Gaia parallaxes will place the calibration on a much improved footing.
There are many RR Lyraes with $6<V<12$ mag and distances less than about 2\,kpc.
Thus Gaia parallaxes should provide excellent calibrations for both $V$ and $K_S$
relations including estimates of their intrinsic scatter and any metallicity
effect on the $K_S$ relation. This will, however, require good infrared
light-curves for the calibrators. In addition good values of [Fe/H] will be required,
either from Gaia itself or from the ground.

 Provided any metallicity effect on the $K_S -\log P$ relation is determined (using
Gaia data) and the low scatter in this relation, just mentioned, is confirmed, this
relation should give better distances than the Gaia parallaxes themselves for RR Lyraes
further away than about 2\,kpc. It will then be possible to carry out extensive 
studies of the distribution (and kinematics) of RR Lyraes over a large volume
(note that an RR Lyrae with a period of 0.5 days has $K_{0} \sim 14.5$\,mag at
10\,kpc.)
This will require extensive $J(H)K_S$ work over the whole sky, but this seems entirely feasible.

\begin{figure}
\begin{center}
 \includegraphics[width=3in]{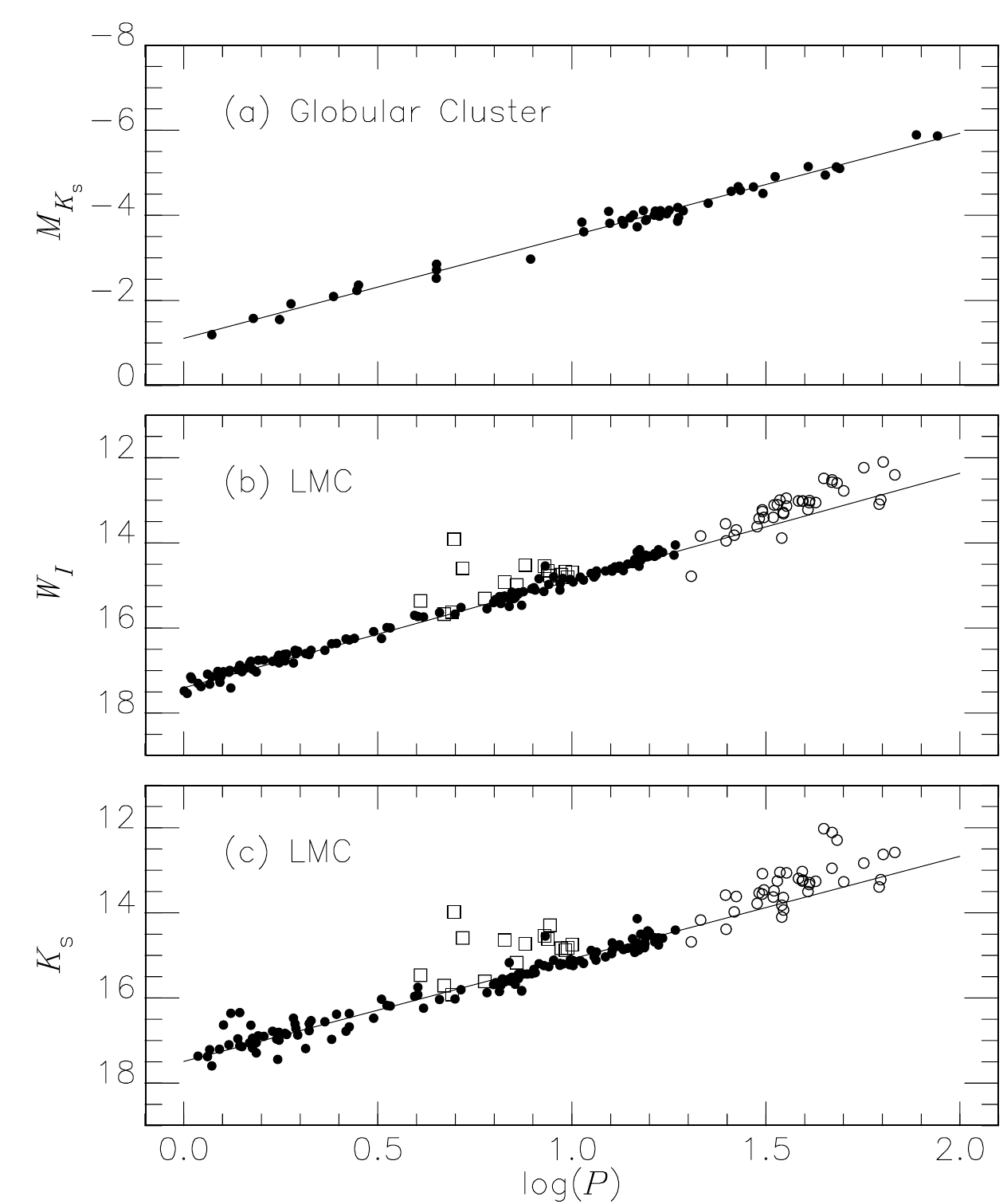} 
 \caption{PL relations for type II Cepheids (see text for details) from Feast (2013).}
   \label{typeII_PL}
\end{center}
\end{figure}

\section{Type-II Cepheids}
A brief mention should be made of the type II Cepheids.
These lie in the same period range as classical Cepheids, but
are low-mass objects belonging to the halo, including globular clusters, 
as well as to the old and/or thick
disc populations. They are often divided into groups according to period:
BL~Her stars ($P<4$\,days), W~Vir stars ($4<P<20$\,days) and RV~Tau stars ($P>20$\,days).
Their potential importance as distance indicators has only recently been fully
recognized. This is due primarily to the work of Matsunaga et al.\ (2006) on type II Cepheids in globular clusters and the LMC work of the OGLE group (e.g. Soszy\'{n}ski et al.\ 2008).
In globular clusters there is a good $K_S -\log P$ relation (see Fig.\,\ref{typeII_PL}a), which may 
extend to the RR Lyrae stars. (e.g.\ Matsunaga et al.\ 2006; Benedict et al. 2011). 
The LMC results (in $K_S$ and $W_{VI}$, Fig.\,\ref{typeII_PL} b and c) are broadly in agreement with  the globular clusters and the low scatter suggests that any metallicity effect is small. However, the LMC work 
reveals a number of complications. In the mean the longer period LMC stars lie above a linear PL relation and there are a significant number
of stars above the PL relations at intermediate periods. These ``peculiar W~Vir"
stars are binaries and can be recognized by their light curves. It is of some interest that, whilst binaries among Classical Cepheids are known in both
our Galaxy and the LMC, there are only a few examples of LMC classical Cepheids
deviating significantly from PL relations (e.g. Soszy\'{n}ski et al.\ 2008).

At present the zero-points of Type II Cepheid relations have to be estimated assuming
the distances to globular clusters or to the LMC, or using pulsation parallaxes. Gaia
should allow direct calibration from parallaxes. 

\section{Mira Variables}
 This and the following sections discuss Mira variables, which are large amplitude pulsators at the tip of the AGB. The  class contains both O-rich and C-rich stars, the latter due to stellar dredge-up processes.  There is currently a great deal of interest in these stars and Gaia should enable major advances in understanding them and in using them to study Galactic structure. 
   
Much of the current interest is related to new mid- and far-infrared observations (e.g. using the Herschel and Spitzer spacecraft) and to the study of mass loss and dust formation which are not quantitatively understood. This is crucial for our understanding of stellar evolution and the composition of the interstellar medium, especially at early times. At the same time, interferometry is providing new insights on the
diameters and overall shapes of these stars as well as on their surface non-uniformities.

\begin{figure}
\begin{center}
 \includegraphics[width=3.4in]{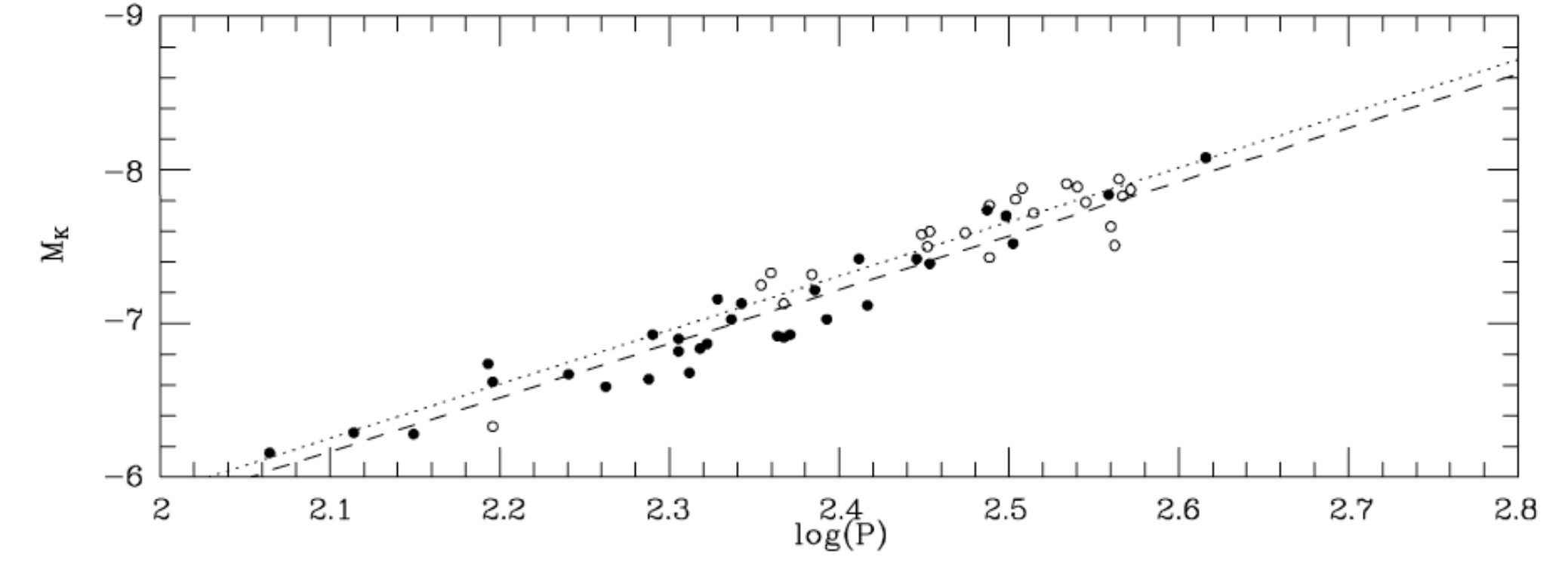} 
 \caption{PL($K$) relation for LMC Miras from Whitelock et al.\ (2008). Open and closed symbols represent O- and C-rich Miras, respectively.}
   \label{LMC_Miras}
\end{center}
\end{figure}

It is known from the LMC that both O- and C-rich Miras follow similar infrared PL relations 
(Fig.\,\ref{LMC_Miras}). These relations show a scatter (per star)
of  0.14\,mag which is clearly an upper limit to the intrinsic scatter. Using Hipparcos parallaxes alone the zero-point for the O-rich stars is obtained with an uncertainty of
$\pm 0.10$\,mag, while including other methods (Whitelock et al.\ 2008) brings this down to $\pm 0.07$\,mag. Since there are large numbers of Galactic Miras (the GCVS gives a very incomplete list of about 8000)  Gaia should in principle make a significant
improvement on this calibration. However, these stars present some serious, and unique, 
challenges to Gaia, as outlined below.

\subsection{The problem of Mira amplitudes and periods}
All Miras have large ranges in visual magnitude; they are defined as having
photographic or $V$ changes in excess of $2.5$\,mag 
(though some stars with smaller amplitudes are similar to Miras)
and the range can exceed 10\,mag for very red stars. Mira itself, $o$ Ceti,
has a range of over 6 mag (e.g. Barth\`es \& Mattei 1997). This will mean that some of the variables
will be either too bright or too faint to be accurately observed by Gaia at some
phases. For instance, whilst the GCVS lists 1883 Miras which are brighter 
than visual or photographic magnitude 12 at maximum, only 18 lie
in the range 6 to 12 mag at all phases, the range for the most accurate
parallaxes (see Table~1). Of these, eleven are O-rich and seven C-rich and
have periods in the range 200 to 450 days.  Additionally, whilst the periods of Miras are typically in the range
100 to 1000 days, the period distribution in the solar neighbourhood peaks 
between 200 and 400 days and there are many Miras with periods near one year which will also present  problems given the baseline for parallax measurement. Hopefully, observations over the life of the Gaia mission will overcome or at least ameliorate these problems. 

\subsection{The problem of Mira sizes}
\begin{figure}
\begin{center}
 \includegraphics[width=2.3in]{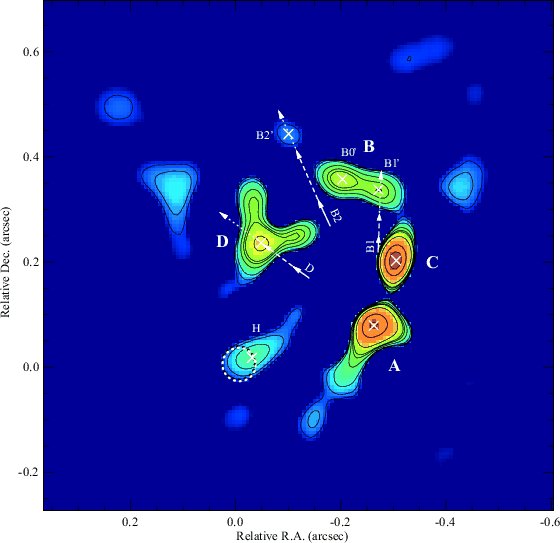} 
 \caption{$H$-band image of the C-rich Mira CW Leo (IRC+10216) from Le\~ao et al.\ (2006), showing only the highest spatial frequencies of the brightness.}
   \label{Leao}
\end{center}
\end{figure}

It has long been known that  the angular sizes of Miras are large and depend on the wavelength
at which they are measured. This is due to the great depth of the atmosphere and the variation of opacity with wavelength. For instance speckle interferometry showed that for O Miras, the diameter in the TiO bands was double that in the nearby continuum  
(Labeyrie et al.\ 1977). The size is a minimum in the near infrared, but even there 
it will be greater than the parallax, e.g., for $o$ Ceti, the Hipparcos parallax is 
$9.3\pm1$\,mas and the angular diameter at $2.2\mu$m$(K)$ is $24$\,mas (Mennesson et al. 2002). 
Early interferometric work (e.g. Lattanzi et al. 1997) showed that these stars are not well 
represented by uniform circular discs. In the case of C Miras, near-infrared photometry
(Whitelock et al.\ 1997, 2006, Feast et al.\ 2003) provided evidence that some
of these stars produce dust erratically as puffs in random directions, after
the manner of R~CrB stars and possibly connected with
large convection cells. Advances in interferometric techniques and high resolution imaging 
(e.g.\  Fig.\,\ref{Leao}) have confirmed this. 
It seems that, at least for dusty Miras, there may be no clear centre of light and that bright 
(low circumstellar absorption) patches may move around in ways not simply related to the
pulsation.

 It is clear that these effects as well as those mentioned in the last section
will complicate the analysis of the astrometry. However, given the wealth of
the expected Gaia data, we may hope that it will offer new insights into
mass loss and convection in these stars, which are at present very poorly 
understood, as well  as giving us valuable parallaxes, proper motions
and radial velocities.

\section {Miras as Age Indicators}
It has been known for some while now that the period of a Mira is a measure of its age,
the relation being similar for O- and C-rich Miras. 
The main evidence for this is the relation between kinematics and period, as well as the
Miras in Galactic globular clusters and in intermediate age
Magellanic Cloud clusters (see Feast (2009) for a summary). We can estimate rough   
ages of $\sim 12$\,Gyr for 200 day Miras and $\sim 2$\,Gyr at 500 days. The
very long period OH/IR Miras must be even younger and more massive.

An interesting case in which this age/mass and period relation appears to be violated
is that of the 551-day Mira in the centre of the globular cluster Lyng\aa\,7.  This object
which is known to be C-rich (in contrast to Miras in other Galactic globular clusters) was
investigated by Feast et al.\ (2013) who established that it was a radial velocity member. Its 
initial mass is expected to have been $\sim 1.5\, M_{\odot}$, i.e, about twice 
the turn-off mass in the cluster. We suggested that it is the evolutionary
product of a stellar merger.  At the cluster distance we find that it has $M_{bol}= -5.0$\,mag,
in excellent agreement with the predictions from the PL relation of $M_{bol} \sim-5.2$\,mag. This is 
particularly interesting as the first example of the PL relation applying to star formed by merging two others. Such mergers 
are likely to be important in dense environments, such as the bulge (see below).
       
\section{Miras and Galactic Structure}
Miras have an important role to play in understanding the Galactic bulge and other
Galactic structures. Early evidence for a bar-like structure in the bulge, with hints of complex underlying structure, was provided by the distribution
of Miras on either side of the Galactic Centre (Whitelock \& Catchpole 1992).
Large scale, near-infrared studies of the bulge region should enable this work to be carried much further.

\begin{figure}
\begin{center}
 \includegraphics[width=2.5in]{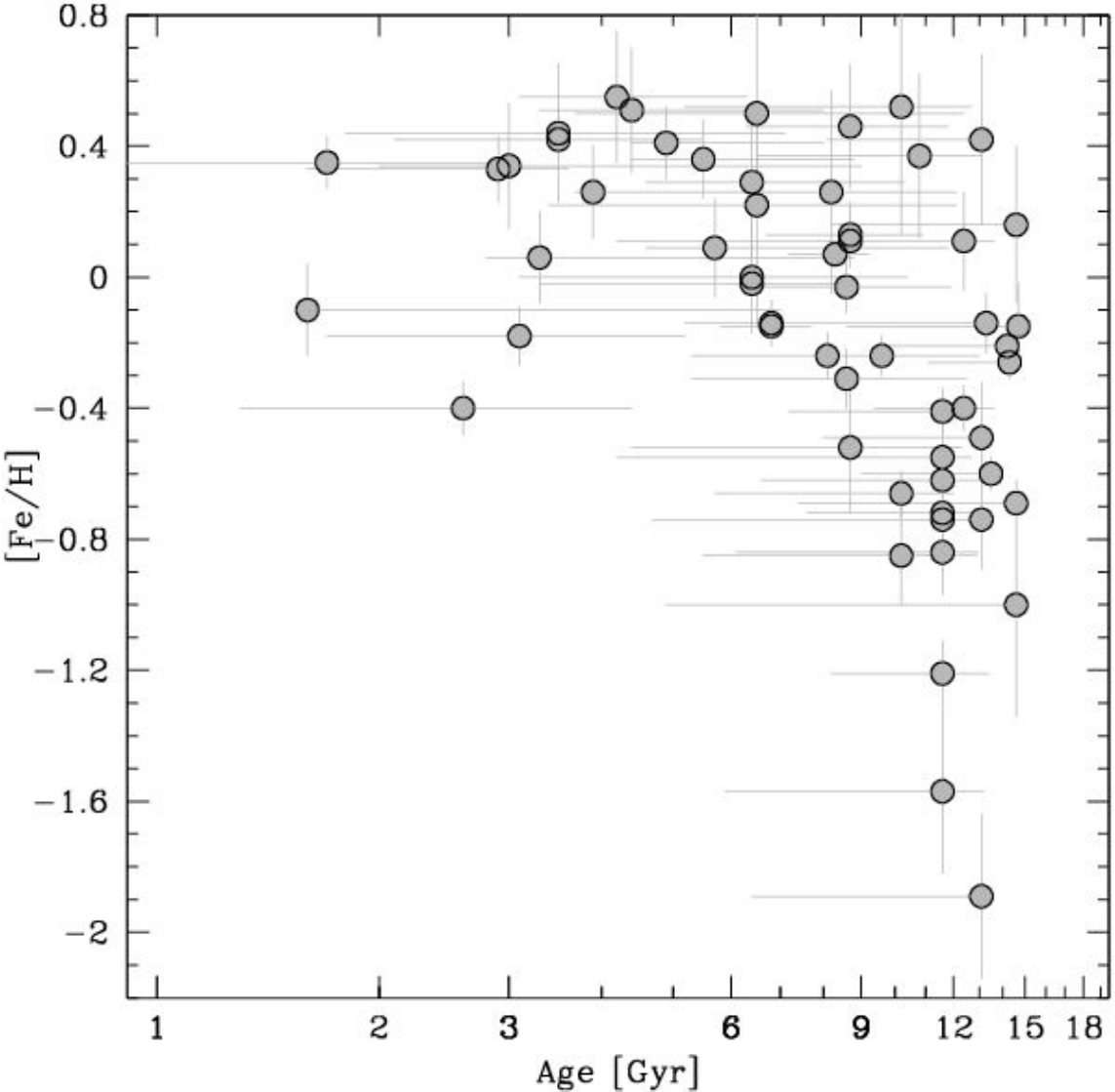} 
 \caption{[Fe/H] as a function of age for dwarfs in the Galactic bulge (Bensby et al.\ 2013) illustrating the
presence of a young population.}
   \label{Bensby}
\end{center}
\end{figure}

As regards the ages of stars in the bulge the Miras have long presented a puzzle.
On the basis of colour-magnitude diagrams it has been claimed that the bulge population, excluding the central regions,  is very old (see e.g. Rich 2013 and references therein). 
However, the bulge contains O Miras with a large range of periods, up to more than 700 days (Whitelock et al. 1991), suggesting a significant intermediate age population.
It was suggested long ago (Greggio \& Renzini 1990) that these long period Miras might
be formed by mergers. The discovery of blue stragglers in the bulge
and the suggestion that they represent an important population there
(e.g. Clarkson et al.\ 2011) makes this suggestion more plausible. However, it         
has also been recently suggested that the bulge does contain an intermediate age and 
relatively metal-rich population (Bensby et al.\ 2013, see Fig.\,\ref{Bensby}). Evidently
the question of the origin of long period Miras in the bulge needs a quantitative
discussion based on much more data. Note that the results discussed in section 7 suggest that Miras may remain good distance
indicators even if they result from mergers.

\begin{figure}
\begin{center}
 \includegraphics[width=1.6in]{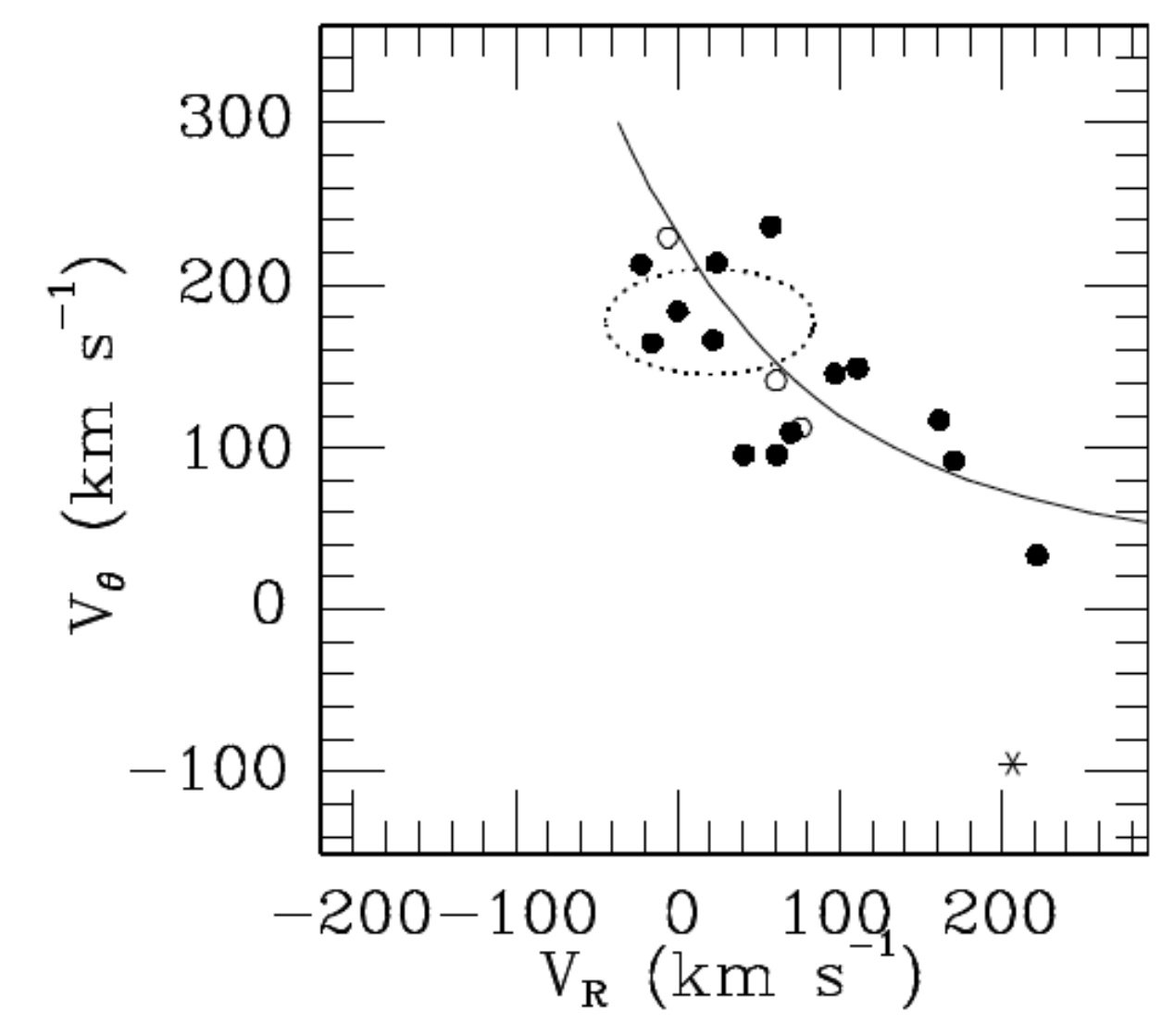} 
 \caption{Space motions of Miras in the Solar neighbourhood with $145<P<200$, illustrating the possible influence of the bar (Feast 2003). }
   \label{Bar}
\end{center}
\end{figure}

Analysis of the space motions of the short period (145-200\,day) Miras (i.e.
a globular cluster-age population) in the general Solar neighbourhood (Feast \& Whitelock 2000) gave the result shown
in Fig.\,\ref{Bar}. This shows that all these Miras with velocities in the
direction of Galactic rotation $V_{\theta} \lesssim 160\,\rm {km\,s^{-1}}$ have positive
(outward) velocities ($V_{R}$) in the Galaxy. It was suggested at the time that this
was related to a bar-like structure. Gaia together with extensive near-infrared
photometry will allow the Galactic kinematics
of Miras of all periods to be studied in detail out to large distances. It will then
be interesting to see whether the hint of structure given by Fig.\,\ref{Bar} is confirmed
and whether it can be fitted with the more complex structure of the bulge, perhaps
an X-shape, which is now being proposed (V\'{a}squez et al.\ 2013) and which
the Mira data also hinted at (Whitelock \& Catchpole 1992).

\section{Miras and the Extragalactic Distance Scale}
  The importance of classical Cepheids for the distance scale is well known
and in recent times the value of these distance indicators in the near- and
the mid-infrared has been stressed in view of the expected capabilities of 
JWST.
The usefulness of Miras at infrared wavelengths is less well known. However,
these stars are bright at infrared wavelengths:\\
At $K$ ($2.2 \mu m$):\\
A 50 day Cepheid has $M_{K} \sim -7.9$ mag\\
A 380 day Mira has $M_{K} \sim -7.9$ mag\\
At $8\mu m$:\\
A 50 day Cepheid has $M_{8} \sim -8.3$ mag\\
A 230 day Mira has $M_{8} \sim -8.3$ mag\\
A 380 day Mira has $M_{8} \sim -9.2$ mag\\
Miras have the additional importance of having a range of ages. They are thus
observable in populations devoid of young stars, e.g. elliptical galaxies, and since they will be less concentrated to the planes of galaxies they will be easier to resolve at large distances.

\begin{figure}%
\centering
\vspace*{-0.7in}
\parbox{2in}{\includegraphics[width=2.2in]{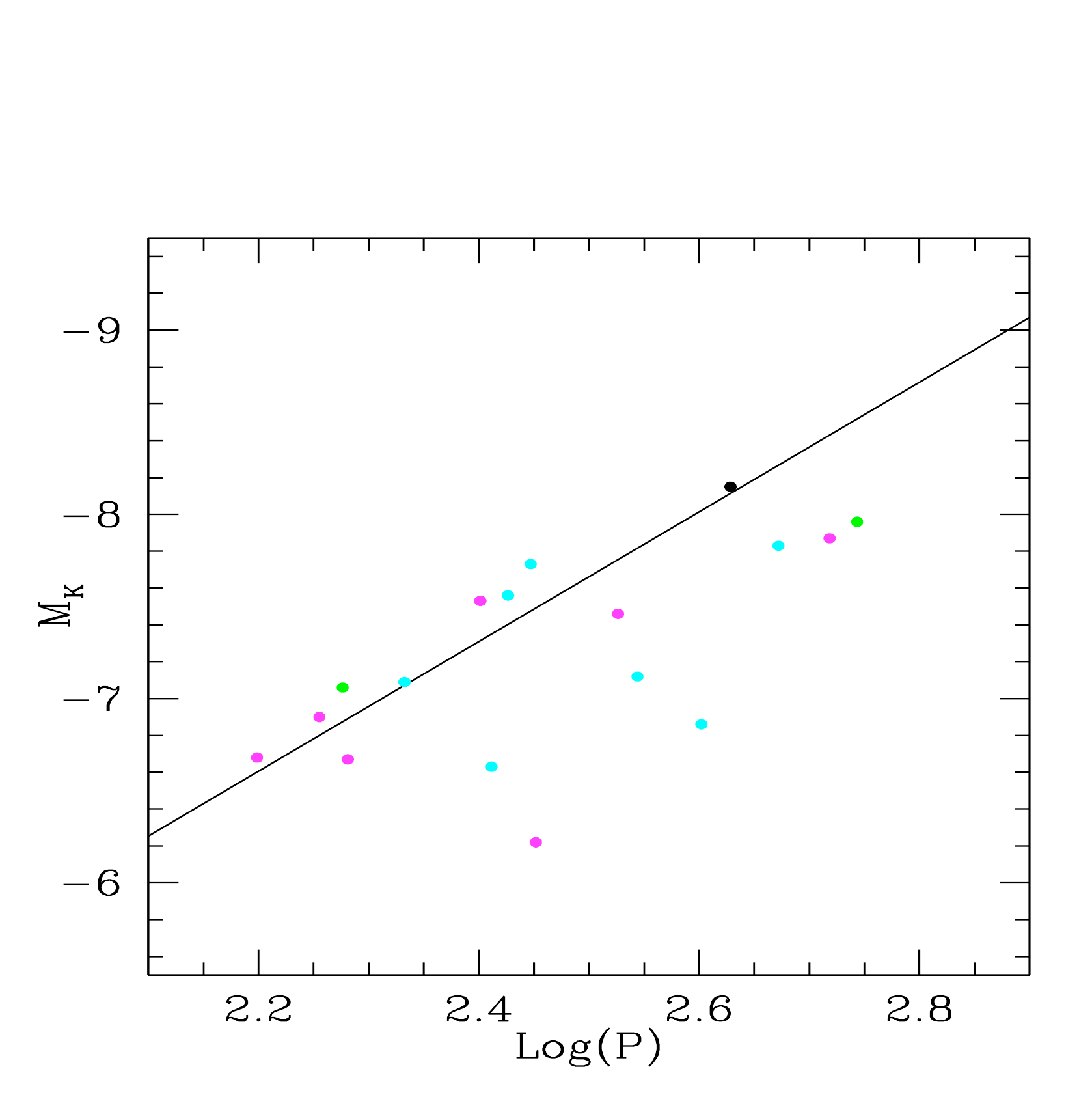} }%
\qquad
\begin{minipage}{2in}%
\vspace*{0.3in}
\includegraphics[width=2.3in]{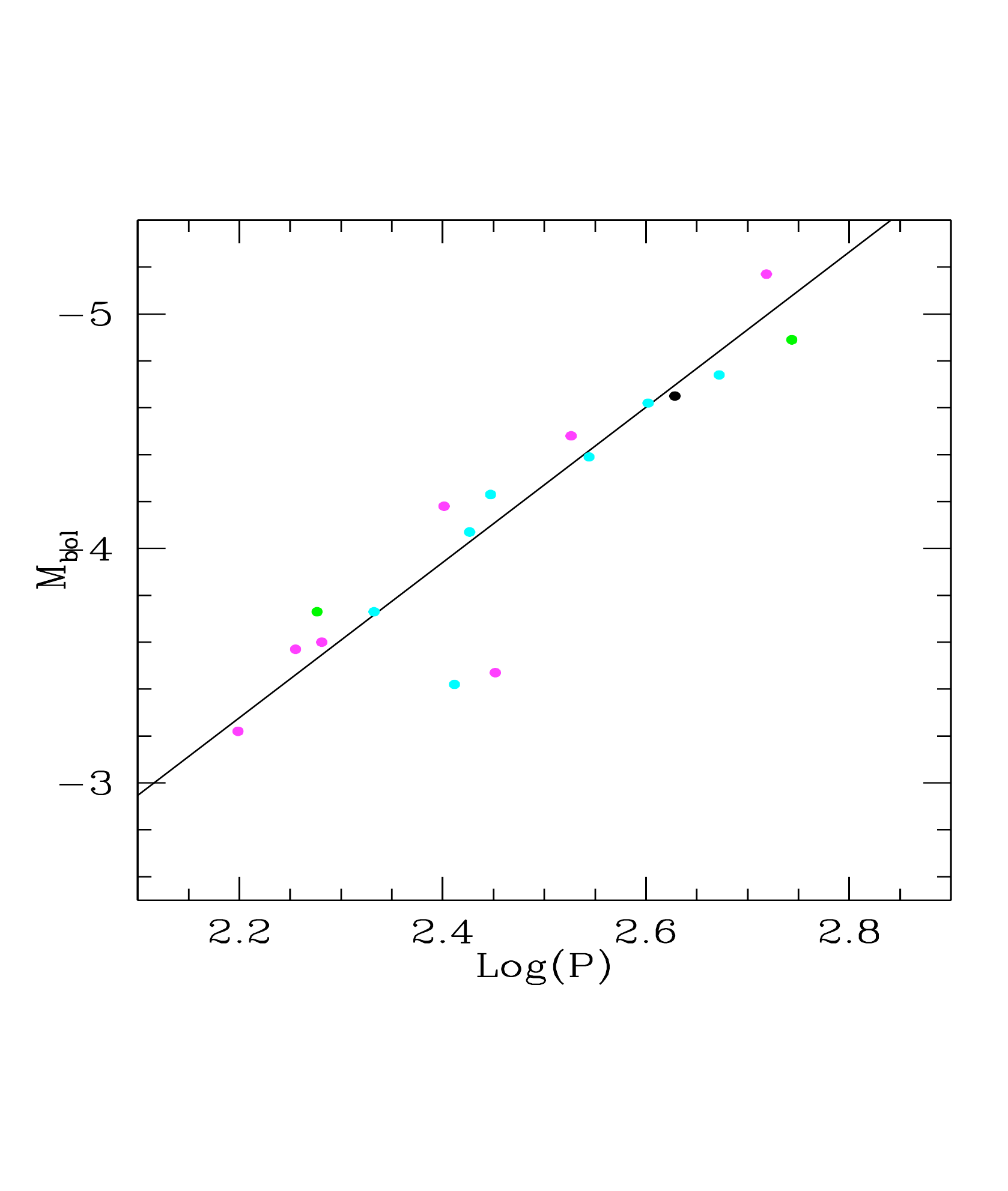} 
\end{minipage}%
\vspace*{-0.2in}
\caption{The PL relations for Local Group dwarf spheroidal galaxies, symbols as follows: Fornax-cyan; Leo~I-magenta; Sculptor-green; Phoenix-black.}%
\label{dsph}%
\end{figure}

\begin{figure}%
\centering
\parbox{2in}{\includegraphics[width=2.0in]{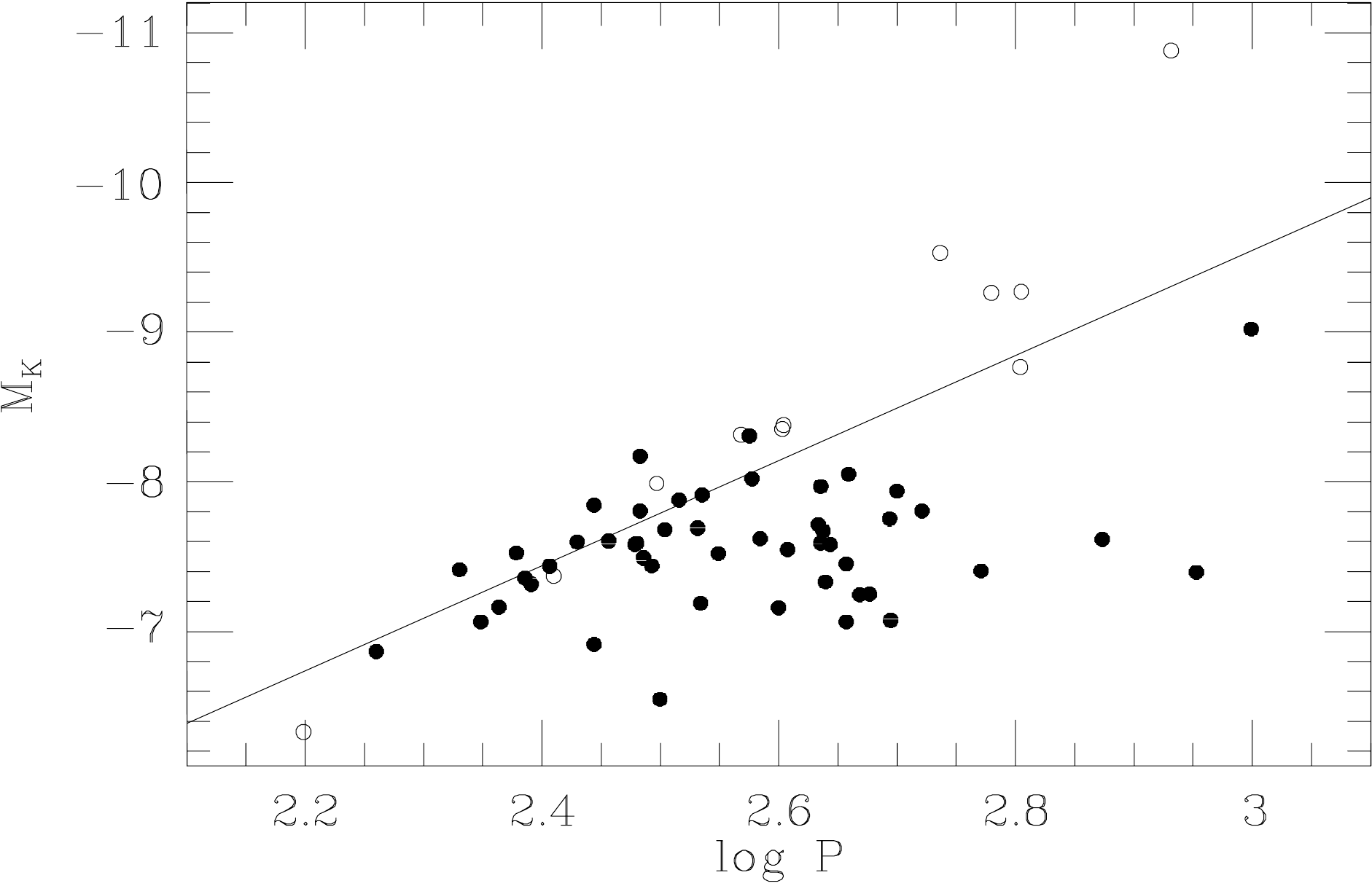} }%
\qquad
\begin{minipage}{2in}%
\includegraphics[width=2.0in]{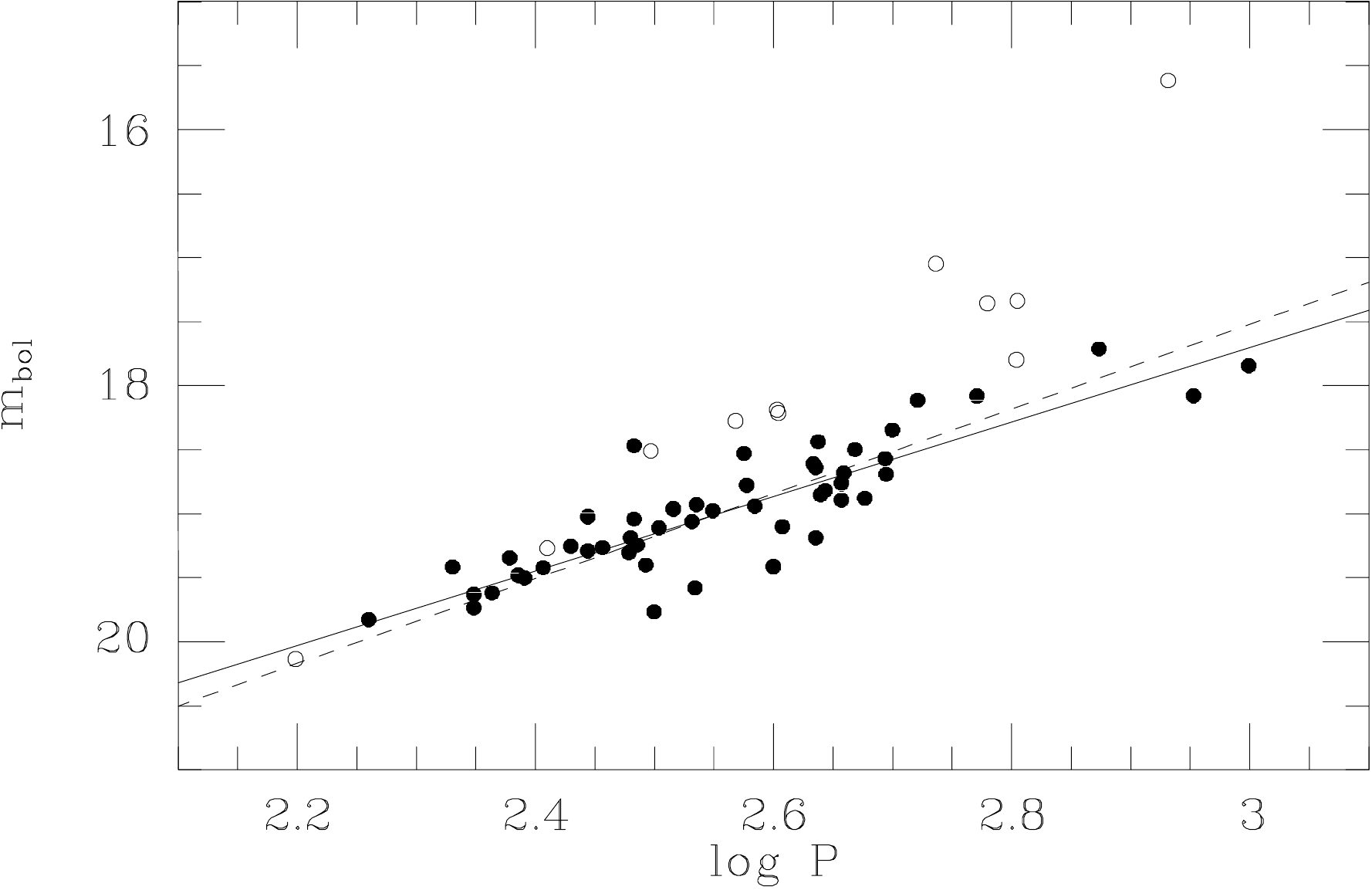} 
\end{minipage}%
\vspace*{0.1in}
\caption{The PL relations for NGC6822 from Whitelock et al.\ (2013). Open and closed circles are O-rich and C-rich Miras, respectively. In the right hand figure the dashed line has the slope of the LMC PL relation, while the solid line was derived from the C stars in NGC\,6822.}%
\label{ngc6822}%
\end{figure}

Figs.~\ref{dsph} and \ref{ngc6822} illustrate the use of Miras as distance indicators in the Local
Group. Fig~\ref{dsph}(a) shows a plot of $M_{K}$ for Miras in a number of Local Group 
dwarf spheroids whose distances were derived by other methods (see Menzies et al. (2008, 2010, 2011) and Whitelock et al. (2009)).
These Miras are all, to the best of our knowledge, carbon-rich. As will be seen there is considerable scatter, 
many stars falling below the line derived from LMC C Miras as a consequence of circumstellar extinction. However, as Fig.\,\ref{dsph}(b) shows, 
a much clearer relation is obtained if one uses bolometric magnitudes. The bolometric
corrections (BC)  were derived from a colour-BC relation. As will be seen, two stars
still fall below the LMC relation. This is due either to the stars undergoing an
RCB-like obscuration event (see above) or to the derived bolometric corrections
being incorrectly estimated for these highly reddened short-period C Miras which 
appear to be unique to the dwarf Spheriodals. The latter idea is supported by the much
brighter bolometric magnitude derived for the variable in Fornax by Sloan et al. (2012) using Spitzer 
data.

Fig.\,\ref{ngc6822}  shows the Mira PL relations in the
the Local Group dwarf irregular NGC6822 (Whitelock et al. 2013). Most of the Miras 
are carbon-rich and it can be seen that the derived bolometric magnitudes
fit a PL relation much better than do the $K$ magnitudes. A few O-rich
Miras are also plotted. At the long periods these tend to lie above the
PL relation both at $K$ and $m_{bol}$. These may be hot bottom burning
stars and/or overtone pulsators.

It will be clear from the above that in using Miras as distance indicators there are a number of possible problems to be kept in mind. As for other distance indicators,
including Cepheids, single objects may give deviant results.

\section{Summary} 

 The discussion of sections 1 to 5 suggests that once firmly calibrated by Gaia,
near-infrared period-luminosity relations of Classical and Type II Cepheids and
RR Lyrae variables can be used to
extend understanding of our own
and nearby galaxies, far beyond the reach of Gaia itself. 

There can be no doubt that Gaia observations of Mira variables will provide
tremendous new insights into the kinematics of different Galactic populations, 
particularly if we can separate these stars according to period. The 
calibration of Mira PL relations will be important for both Galactic
and extragalactic work. It also seems likely that detailed Gaia observations of a few nearby Miras together with ground based observations, including interferometry, will revolutionize our understanding of Mira surface structure, of what is actually
happening during pulsation cycles and of how mass loss is driven. 

However, to fully exploit the Gaia results 
for all the classes of variables discussed in this review, extensive near-infrared photometry (e.g. $JHK_S$ or similar) will be required.

\acknowledgements
 The authors gratefully
acknowledge the receipt of research grants from the National Research 
Foundation (NRF) of South Africa and thank John Menzies for his corrections.

\begin{discussion}

\discuss{Ivezi\'{c}}
{Could you please summarize physical reasons for the age-period
relation for Miras, and comment on whether this relation depends on metallicity?}
\discuss{Whitelock}
{It is difficult to answer this fully because
all theoretical work depends on assumptions about mass loss at the tip of the AGB where the Miras lie. The observations we discussed indicate that such a
relation exists though we do not have a very clear idea of its scatter, i.e. the
range of periods at a given age (and metallicity). As regards metallicity dependence,
we know that in globular clusters the Mira period is larger for more metal-rich clusters 
(e.g. Feast \& Whitelock 2000). This might for instance be due to atmospheric structure affecting the stellar radius, but that is not clear. Further work on the age-period relation is desirable, especially for C-Miras. However, C and O-Miras
have rather similar absolute magnitudes at the same period and probably
similar ages.} 
\discuss{Lecai Deng} {Are those outliers in the two PLs the same? Is there any physical reason for them to be outliers?}
\discuss{Whitelock} {They are either undergoing line-of-site obscuration (see section 6.2) or the bolometric correction does not apply to these short period dusty stars, which seem unique to the dwarf speheroidals.}

\end{discussion}

\end{document}